\documentclass[longauth]{aa} 

\makeatletter
\renewcommand*{\@fnsymbol}[1]{\ifcase#1\or*\or$\dagger$\or$\ddagger$\or**\or$\dagger\dagger$\or$\ddagger\ddagger$\fi}
\makeatother

\usepackage{natbib}
\bibpunct{(}{)}{;}{a}{}{,}

\usepackage{graphicx}
\usepackage{txfonts}
\usepackage{xspace}

\usepackage{url,hyperref}

\usepackage{footmisc}

\usepackage{rotating}

\usepackage[switch]{lineno} 

\newcommand{\eboss}{E-BOSS catalogue\xspace}  
\newcommand{\seboss}{second E-BOSS catalogue release\xspace}  
\newcommand{\hess}{H.E.S.S.\xspace}

\newcommand{\zetaoph}{$\zeta$~Ophiuchi\xspace}
\newcommand{\bd}{BD$+43\degr3654$\xspace}

\begin{document} 

\title{Systematic search for very-high-energy gamma-ray emission from bow shocks of runaway stars} 
\titlerunning{Search for very-high-energy stellar bow shock emission}
\authorrunning{The H.E.S.S. Collaboration}

\author{H.E.S.S. Collaboration
\and H.~Abdalla \inst{1}
\and A.~Abramowski \inst{2}
\and F.~Aharonian \inst{3,4,5}
\and F.~Ait Benkhali \inst{3}
\and A.G.~Akhperjanian\protect\footnotemark[2]$^\dagger$\inst{6,5} 
\and T.~Andersson \inst{10}
\and E.O.~Ang\"uner \inst{21}
\and M.~Arakawa \inst{43}
\and M.~Arrieta \inst{15}
% ...
}

\institute{
Centre for Space Research, North-West University, Potchefstroom 2520, South Africa \and 
Universit\"at Hamburg, Institut f\"ur Experimentalphysik, Luruper Chaussee 149, D 22761 Hamburg, Germany \and 
Max-Planck-Institut f\"ur Kernphysik, P.O. Box 103980, D 69029 Heidelberg, Germany \and 
Dublin Institute for Advanced Studies, 31 Fitzwilliam Place, Dublin 2, Ireland \and 
National Academy of Sciences of the Republic of Armenia,  Marshall Baghramian Avenue, 24, 0019 Yerevan, Republic of Armenia  \and
Yerevan Physics Institute, 2 Alikhanian Brothers St., 375036 Yerevan, Armenia \and
Institut f\"ur Physik, Humboldt-Universit\"at zu Berlin, Newtonstr. 15, D 12489 Berlin, Germany \and
University of Namibia, Department of Physics, Private Bag 13301, Windhoek, Namibia \and
GRAPPA, Anton Pannekoek Institute for Astronomy, University of Amsterdam,  Science Park 904, 1098 XH Amsterdam, The Netherlands \and
Department of Physics and Electrical Engineering, Linnaeus University,  351 95 V\"axj\"o, Sweden \and
Institut f\"ur Theoretische Physik, Lehrstuhl IV: Weltraum und Astrophysik, Ruhr-Universit\"at Bochum, D 44780 Bochum, Germany \and
GRAPPA, Anton Pannekoek Institute for Astronomy and Institute of High-Energy Physics, University of Amsterdam,  Science Park 904, 1098 XH Amsterdam, The Netherlands \and
Institut f\"ur Astro- und Teilchenphysik, Leopold-Franzens-Universit\"at Innsbruck, A-6020 Innsbruck, Austria \and
School of Physical Sciences, University of Adelaide, Adelaide 5005, Australia \and
LUTH, Observatoire de Paris, PSL Research University, CNRS, Universit\'e Paris Diderot, 5 Place Jules Janssen, 92190 Meudon, France \and
Sorbonne Universit\'es, UPMC Universit\'e Paris 06, Universit\'e Paris Diderot, Sorbonne Paris Cit\'e, CNRS, Laboratoire de Physique Nucl\'eaire et de Hautes Energies (LPNHE), 4 place Jussieu, F-75252, Paris Cedex 5, France \and
Laboratoire Univers et Particules de Montpellier, Universit\'e Montpellier, CNRS/IN2P3,  CC 72, Place Eug\`ene Bataillon, F-34095 Montpellier Cedex 5, France \and
DSM/Irfu, CEA Saclay, F-91191 Gif-Sur-Yvette Cedex, France \and
Astronomical Observatory, The University of Warsaw, Al. Ujazdowskie 4, 00-478 Warsaw, Poland \and
Aix Marseille Universit\'e, CNRS/IN2P3, CPPM UMR 7346,  13288 Marseille, France \and
Instytut Fizyki J\c{a}drowej PAN, ul. Radzikowskiego 152, 31-342 Krak{\'o}w, Poland \and
Funded by EU FP7 Marie Curie, grant agreement No. PIEF-GA-2012-332350,  \and
School of Physics, University of the Witwatersrand, 1 Jan Smuts Avenue, Braamfontein, Johannesburg, 2050 South Africa \and
Laboratoire d'Annecy-le-Vieux de Physique des Particules, Universit\'{e} Savoie Mont-Blanc, CNRS/IN2P3, F-74941 Annecy-le-Vieux, France \and
Landessternwarte, Universit\"at Heidelberg, K\"onigstuhl, D 69117 Heidelberg, Germany \and
Universit\'e Bordeaux, CNRS/IN2P3, Centre d'\'Etudes Nucl\'eaires de Bordeaux Gradignan, 33175 Gradignan, France \and
Oskar Klein Centre, Department of Physics, Stockholm University, Albanova University Center, SE-10691 Stockholm, Sweden \and
Wallenberg Academy Fellow,  \and
Institut f\"ur Astronomie und Astrophysik, Universit\"at T\"ubingen, Sand 1, D 72076 T\"ubingen, Germany \and
Laboratoire Leprince-Ringuet, Ecole Polytechnique, CNRS/IN2P3, F-91128 Palaiseau, France \and
APC, AstroParticule et Cosmologie, Universit\'{e} Paris Diderot, CNRS/IN2P3, CEA/Irfu, Observatoire de Paris, Sorbonne Paris Cit\'{e}, 10, rue Alice Domon et L\'{e}onie Duquet, 75205 Paris Cedex 13, France \and
Univ. Grenoble Alpes, IPAG,  F-38000 Grenoble, France \protect\\ CNRS, IPAG, F-38000 Grenoble, France \and
Department of Physics and Astronomy, The University of Leicester, University Road, Leicester, LE1 7RH, United Kingdom \and
Nicolaus Copernicus Astronomical Center, Polish Academy of Sciences, ul. Bartycka 18, 00-716 Warsaw, Poland \and
Institut f\"ur Physik und Astronomie, Universit\"at Potsdam,  Karl-Liebknecht-Strasse 24/25, D 14476 Potsdam, Germany \and
Friedrich-Alexander-Universit\"at Erlangen-N\"urnberg, Erlangen Centre for Astroparticle Physics, Erwin-Rommel-Str. 1, D 91058 Erlangen, Germany \and
DESY, D-15738 Zeuthen, Germany \and
Obserwatorium Astronomiczne, Uniwersytet Jagiello{\'n}ski, ul. Orla 171, 30-244 Krak{\'o}w, Poland \and
Centre for Astronomy, Faculty of Physics, Astronomy and Informatics, Nicolaus Copernicus University,  Grudziadzka 5, 87-100 Torun, Poland \and
Department of Physics, University of the Free State,  PO Box 339, Bloemfontein 9300, South Africa \and
Heisenberg Fellow (DFG), ITA Universit\"at Heidelberg, Germany  \and
GRAPPA, Institute of High-Energy Physics, University of Amsterdam,  Science Park 904, 1098 XH Amsterdam, The Netherlands \and
Department of Physics, Rikkyo University, 3-34-1 Nishi-Ikebukuro, Toshima-ku, Tokyo 171-8501, Japan
% ...
}

\offprints{H.E.S.S.~collaboration,
\protect\\\email{\href{mailto:contact.hess@hess-experiment.eu}{contact.hess@hess-experiment.eu}};
\protect\\\protect\footnotemark[1] $*$ Corresponding authors
\protect\\\protect\footnotemark[2] $\dagger$ Deceased
}

\makeatletter
\renewcommand*{\@fnsymbol}[1]{\ifcase#1\@arabic{#1}\fi}
\makeatother

   \date{Preprint online version: April 24,2017 }

 \abstract
   {Runaway stars form bow shocks by ploughing through the interstellar medium at supersonic speeds and are promising sources of non-thermal emission of photons. One of these objects has been found to emit non-thermal radiation in the radio band. This triggered the development of theoretical models predicting non-thermal photons from radio up to very-high-energy (VHE, E $\geq 0.1$\,TeV) gamma rays. Subsequently, one bow shock was also detected in X-ray observations. However, the data did not allow discrimination between a hot thermal and a non-thermal origin. Further observations of different candidates at X-ray energies showed no evidence for emission at the position of the bow shocks either. A systematic search in the \textit{Fermi}-LAT energy regime resulted in flux upper limits for 27 candidates listed in the \eboss. }
   {Here we perform the first systematic search for VHE gamma-ray emission from bow shocks of runaway stars.}
   {Using all available archival \hess data we search for very-high-energy gamma-ray emission at the positions of bow shock candidates listed in the \seboss. Out of the 73 bow shock candidates in this catalogue, 32 have been observed with \hess }
   {None of the observed 32 bow shock candidates in this population study show significant emission in the \hess energy range. Therefore, flux upper limits are calculated in five energy bins and the fraction of the kinetic wind power that is converted into VHE gamma rays is constrained. }
   {Emission from stellar bow shocks is not detected in the energy range between $0.14$ and $18$\,TeV. The resulting upper limits constrain the level of VHE gamma-ray emission from these objects down to 0.1 – 1\% of the kinetic wind energy.}
	\maketitle

\section{Introduction}
Stars with velocities larger than $\sim$30\,km\,s$^{-1}$ (corrected for Solar motion and Galactic rotation) are historically called runaway stars due to their fast movement
away from OB associations. Two scenarios for the formation process of runaway stars have been proposed: 
the dynamical ejection and the binary supernova scenario. \cite{1967BOTT....4...86P} used simulations to verify that during the collapse of a small cluster, dynamical interactions of the stars can lead to high spatial velocities. \cite{1957moas.book.....Z} suggested that the runaway stars are formed during the supernova explosion in a binary system, where the second star keeps its high spatial velocity due to sudden mass loss during the supernova event. \cite{2000ApJ...544L.133H} showed that both proposed mechanisms take place in nature by retracing star trajectories. Examples for the supernova scenario and the dynamical ejection scenario are \zetaoph and AE Aurigae, respectively. 

Since these massive OB stars have very fast stellar winds with velocities up to a few thousand kilometer per second, comparable to the shock
speed of young supernova remnants, they are promising candidates for the acceleration of particles (electrons/protons) to high energies producing non-thermal emission.
Stars moving through the interstellar medium (ISM) at supersonic speeds sweep matter
up in their direction of motion and form bow shocks. The swept-up dust in these large-scale bow shocks is heated and ionized by the stellar
radiation, which leads to infrared emission.
The thermal emission of these coma-shaped features was first discovered by \cite{vanBuren88} using data from the \textit{Infrared Astronomical
Satellite} (\textit{IRAS}). 
The first survey of stellar bow shocks was performed by \cite{vanBuren95}, followed by the Extensive stellar BOw Shock Survey catalogue \citep[E-BOSS;][]{eboss}.

\cite{benaglia_bd} were the first to report on the detection of non-thermal radio emission from a stellar bow shock, namely \bd. They introduced an emission model predicting
non-thermal photons detectable at radio, X-ray and gamma-ray energies. In this model, charged particles are accelerated up to relativistic energies via Fermi
acceleration in the shock wave originating from the supersonic motion of the star. These relativistic particles interact with the ambient matter, photon
or magnetic fields and produce non-thermal emission. The bow shock system is composed of two shocks, a slow forward shock with the ISM and a fast reverse shock with the stellar wind in which the relativistic particles are accelerated more efficiently. A more detailed description of the model and further developments can be found in \cite{benaglia_bd},
\cite{Valle_model_zeta}, \cite{ae_aurigae_LopezSantiago} and \cite{pereira_modelling}. 

Based on this model, several observations of promising bow shock candidates followed, aiming to detect non-thermal emission. The follow-up search by \cite{terada_bd} for a non-thermal X-ray counterpart of \bd using data from \textit{Suzaku} revealed no emission in this regime.
However, the resulting upper limits imply that the emission model from \cite{benaglia_bd} overestimated either the efficiency of the shock-heating process, leading to electron energies that do not exceed 10\,TeV, or the grade of turbulence of the magnetic field in the shock acceleration region. 
Further X-ray observations of \zetaoph and \bd \citep{toala_xray} resulted in upper
limits for non-thermal emission and lead to the conclusion that the intensity of the emission is below the sensitivity of current X-ray satellites. X-ray observations with \textit{XMM-Newton} of AE Aurigae (HIP 24575) revealed for the first time
significant emission, but its nature (very hot thermal or non-thermal) could not be firmly determined \citep{ae_aurigae_LopezSantiago}. In the case of
\zetaoph, \cite{toala_xray} detected diffuse emission in the vicinity of this candidate, which they attribute to a plasma with a temperature of
$2\cdot 10^6$\,K, in agreement with predictions of high plasma temperatures caused by instabilities mixing material between the shocked wind and the
photo-ionized gas at the wake of the shock \citep{Mackey_2015}. Recently, \citep{toala2017} showed that the X-ray emission close to AE Aurigae is point-like and unrelated to the bow shock. They furthermore searched for non-thermal diffuse X-ray emission around 6 well-determined runaway stars and found no evidence for it.   

\cite{Valle_pulsar} suggested the high-energy (HE, 100\,MeV to $\sim$100\,GeV) gamma-ray source 2FGL J2030.7+4417 \citep{2fgl} to be associated with the bow shock of HD 195592. However, the source 2FGL J2030.7+4417 has been identified as a gamma-ray pulsar \citep{Pletsch_pulsar} and shows no significant off-pulse emission \citep{2013ApJS..208...17A}, a strong indication that the detected photons predominantly originate in the pulsar and not in the bow shock.
 
A possibility of stellar bow shocks being variable gamma-ray sources was introduced by \cite{delValle_variable}. The predicted variability in the gamma-ray flux originates from inhomogeneities of the ambient medium, leading to changes in the physical properties and thus the luminosity. The expected time-scale of the variations is $\sim$1 year and depends on the size and density gradient of the molecular cloud and the velocity of the star. 

In the HE gamma-ray regime \cite{Schulz_fermi} performed the first systematic search for non-thermal
emission from bow shocks around runaway stars using 57 months of
\textit{Fermi}-LAT data testing the predictions of \cite{Valle_model_zeta}, \cite{benaglia_bd} and \cite{ae_aurigae_LopezSantiago}. This population study resulted in upper limits for 27 bow shocks including \zetaoph for which the upper limit on its emission
was found to be a factor $\sim$5 below the predicted emission from \cite{Valle_model_zeta}.

In this work, we search for very-high-energy (VHE, E $\geq 0.1$\,TeV) emission from
stellar bow shocks using the latest, most comprehensive survey of bow shocks of runaway stars \citep[][\seboss]{eboss_r2} which uses recent
infrared data releases, mainly from the \textit{Wide-field Infrared Survey Explorer (WISE)}. The \seboss includes bow shocks from literature and
serendipitously found ones to complete the sample. It comprises 73 bow shock candidates: 28 candidates from the first \eboss \citep{eboss}, 16
new ones and 29 from recent publications. 

We describe the \hess observations, data analysis and results of 32 bow shock candidates in Sect.\,2. A discussion on the implications of these non-detections
is presented in Sect.\,3.

\section{Observations, data analysis and results}
\hess is an array of imaging atmospheric Cherenkov telescopes located in the Khomas Highland in Namibia at an altitude of 1800 m above sea level
\citep[$23^{\circ} 16' 18"$ S, $16^{\circ} 30' 00"$ E;][]{hinton_hess_2004}. The initial configuration of four 12 meter telescopes (\hess phase I) was extended
with a central 28 meter telescope in July 2012. This work only uses data from the initial configuration, which provides an energy threshold of
$\sim$100\,GeV with an angular resolution better than 0.1$\degr$ and an energy resolution below 15\%. The standard quality selection was used to
discard observations during bad weather or instrumental conditions \citep{hess_crab_2006}. 

The \seboss \citep{eboss_r2} is the basis for the population study presented in this work. Almost 50$\%$ (32 out of 73) of the candidates in the \seboss are covered by \hess observations. 27 of the observed bow shocks are located within the Galactic plane, profiting from the nine-year-long \hess Galactic Plane Survey \citep[HGPS;][]{hgps} of the inner Milky Way. The study presented here complements the population studies on pulsar wind nebulae \citep{pwnpop} and supernova remnants \citep{snrpop}. 

The coordinates listed in the \seboss are the stars' coordinates for all candidates except the seven serendipitous discoveries "SER1--7" for which the star could not be firmly identified. In these cases, the apex of the bow shock was estimated visually using publicly available \textit{WISE} data. The four corresponding candidates in the \hess sample are marked with *** in Table\,\ref{table:UpperLimits}. 

The \hess analyses are performed for the positions given in the \seboss with seven exceptions:  
The three bow shocks in M 17 have an angular separation of less than $0.1\degr$ which is not resolvable for \hess due to its point spread function (PSF $\sim0.1^\circ$; \cite{hess_crab_2006}). For these three objects one analysis was performed for the position of M~17-S2 which is in the centre of the three. The two exceptions M~17-S1 and S3 are marked with a * in Table\,\ref{table:UpperLimits}, since the coordinates of M~17-S2 are used for the analysis. 

To ensure that the defined source region of the analysis covers the bow shock, its size and distance to the star have to be evaluated. If the length of the bow shock listed in the \seboss is larger than $0.1\degr$ we estimate the bow shock position visually using publicly available \textit{WISE} data and perform the analysis for this position. This criterion leads to updated coordinates for HIP 32067, HIP 88652, HIP 92865, Star 1 and G2; they are marked with ** in Table\,\ref{table:UpperLimits}. For HIP 32067 with a length of 13' ($>2*0.1\degr$) the source region was enlarged from the standard $0.1\degr$ to $0.11\degr$. 

The data were analysed using the \emph{ImPACT} analysis method described in \cite{parsons_impact_2014}. The “standard cuts” of this analysis were adopted,
including a minimum charge of 60 photoelectrons per shower image and a signal extraction region of $0.1\degr$. A cross-check analysis performed with the \emph{model} analysis method as presented by \cite{2009APh....32..231D} yielded compatible results.

The differential upper limits are presented in Table\,\ref{table:UpperLimits}, including the duration of the \hess observations (live-time) and the parameters for each star. To avoid potential systematic biases, upper limits are only calculated if more than 10 events are recorded in the OFF regions that are used to estimate the background.

The analyses of all bow shock candidates were performed in a systematic way by using the same analysis cuts and configurations.
None of the analysed bow shock candidates showed statistically significant VHE gamma-ray emission at the position of the bow shock; thus, upper limits on the flux are calculated using the method presented by \cite{Rolke2005}.

In some cases, the candidates are close to known VHE gamma-ray sources, which leads to significances up to 3$\,\sigma$. However, dedicated analyses of the sky maps and the squared-angular distance distributions of the reconstructed direction of the events with respect to the candidates' source positions clearly showed that the emission is not originating from the bow shock. In these cases upper limits were calculated in the same way as for the rest of the population. 

The differential gamma-ray flux upper limits at $95\%$ confidence-level in five energy bins (equally spaced in logarithmic energy) assuming a power-law spectrum of gamma-ray emission ($d\Phi/dE = \Phi_0 {(E/E_0) }^{-\Gamma}$) with a photon index $\Gamma=2.5$ are presented in Table\,\ref{table:UpperLimits} and in Fig.\,\ref{Fig:ul_models}.
Assuming different indices ($\Gamma=2.0$ and $\Gamma=3.0$) leads to marginal changes in the upper limits of order $10\%$ or less.

\section{Discussion and conclusions}
\label{sec:results}
\begin{figure*}
  \includegraphics[width=\textwidth]{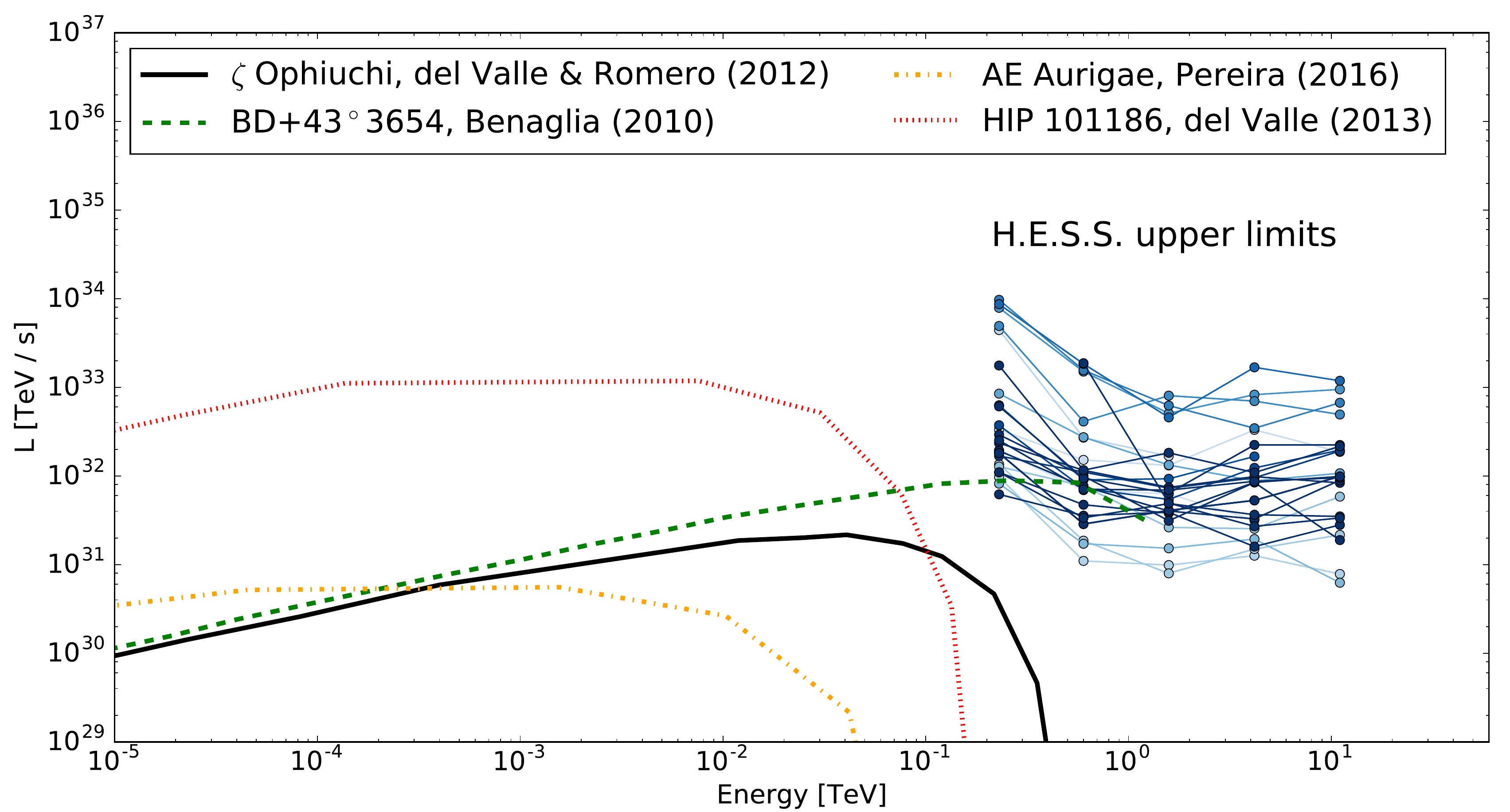}
  \caption{VHE gamma-ray luminosity upper limits for the 28 bow shock candidates with known distance, compared to model predictions for four different bow shocks (none of these four is in the \hess survey sample, see text for details).}
  \label{Fig:ul_models}
\end{figure*}

There are currently no model predictions published for the bow shocks analysed here. Therefore, the published models for four different bow shocks are shown together with the upper limits from this work in Fig.\,\ref{Fig:ul_models}. All four model predictions are based on the model by \cite{Valle_model_zeta} and were motivated by dedicated searches for non-thermal emission 
The comparison of the VHE upper limits with the model prediction for \bd (the only confirmed non-thermal emitter) suggests that several candidates of the \hess bow shock sample do not emit VHE gamma rays at the level predicted by \cite{benaglia_bd}. 

\begin{sidewaystable*}
\small
\rule{0cm}{20cm}   
\caption{Differential gamma-ray flux upper limits ($95\%$ confidence-level) for bow shocks of runaway stars. ID, Star, distance $d$, wind velocity  $v_\mathrm{wind}$ and mass-loss rate $\dot{M}$ as listed in \cite{eboss_r2}. The positions with the coordinates l and b denote the ones used for the analysis, which is not in all cases equal to the one in  \cite{eboss}  and \cite{eboss_r2}  (see text for details). Wind velocity: All values from \cite{eboss}; brackets indicate values adopted from stars with the same spectral type. The live-time, how long each object was observed with \hess, is also listed. The acceptance-corrected live-time (the observation time corrected for the non-uniform acceptance across the field of view of the camera) is given in parentheses.  }              
\label{table:UpperLimits}      
\centering                                        
\begin{tabular}{l|l|r|r|r|r|r||r|r|r|r|r|r||r}          
\hline\hline         
ID & Star & l & b & $d$ & $v_\mathrm{wind}$  & $\dot{M}$  & Live-time &  \multicolumn{6}{c}{Upper limits $E^2$ d$\Phi$ /dE [$10^{-12}\,$TeV\,cm$^{-2}$\,s$^{-1}$]} \\
 & &  &  &  &  &  &  (acc.-corr.) &   \multicolumn{6}{c}{Energy bins [TeV]} \\  
 & & [\degr] &[\degr] & [pc] & [km/s] & $10^{-6}M_\odot $/yr   & [h] & $0.14-0.37$ & $0.37-0.97$  &$0.97-2.57$ & $2.57-6.78$ & $6.78-17.92 $  & $0.1 - 10 $ \\
 \hline 
EB8 & HIP 25923 & 210.44 & $-$20.98 & 900 & [1000] & 0.06 & 3.9 (2.2) & 2.01 & 0.71 & 0.65 & 0.33 & - & 0.40 \\ 
EB13 & HIP 32067** & 206.20 & 0.85 & 2117$\pm$367 & 2960 & 0.13 & 21.7 (9.4) & 0.62 & 0.28 & 0.25 & 0.62 & 0.35 & 0.12 \\ 
EB15 & HIP 38430 & 243.16 & 0.36 & 900 & [2570] & 0.70 & 1.8 (0.1) & 45.73 & 2.79 & 1.73 & - & - & 0.80 \\ 
EB17 & HIP 72510 & 318.77 & 2.77 & 350 & [2545] & 0.27 & 12.8 (2.4) & 6.64 & 0.75 & 0.67 & 0.86 & 0.53 & 0.58 \\ 
EB18 & HIP 75095 & 322.68 & 0.91 & 800 & [1065] & 0.14 & 22.5 (13.9) & 1.77 & 0.24 & 0.10 & 0.20 & 0.28 & 0.15 \\ 
EB23 & HIP 88652** & 15.11 & 3.36 & 650 & [1535] & 0.50 & 9.2 (2.5) & 2.53 & 1.52 & 0.52 & 0.50 & 1.16 & 0.72 \\ 
EB24 & HIP 92865** & 41.75 & 3.41 & 350 & [1755] & 0.04 & 3.9 (2.3) & 5.60 & 1.17 & 1.04 & 1.33 & 0.43 & 0.87 \\ 
EB32 & SER1*** & 264.78 & 1.54 & - & 250 & 0.03 & 3.0 (2.0) & 2.92 & 1.06 & 0.32 & 0.77 & - & 0.59 \\ 
EB33 & HIP 44368 & 263.10 & 3.90 & 1900$\pm200^a$ & 1100 & 0.80 & 7.9 (6.1) & 1.97 & 0.63 & 0.31 & 0.20 & 0.25 & 0.36 \\ 
EB36 & SER2*** & 282.48 & $-$2.46 & - & - & - & 15.7 (7.9) & 1.10 & 0.54 & 0.24 & 0.18 & 0.33 & 0.22 \\ 
EB37 & RCW 49-S1 & 284.08 & 0.43 & 6100 & 2800 & 3.23 & 51.3 (29.2) & 1.77 & 0.34 & 0.11 & 0.19 & 0.21 & 0.17 \\ 
EB38 & RCW 49-S2 & 284.30 & 0.30 & 6100 & 2600 & 0.60 & 51.7 (31.5) & 1.11 & 0.09 & 0.18 & 0.16 & 0.11 & 0.07 \\ 
EB39 & RCW 49-S3 & 284.34 & 0.20 & 6100 & 2800 & 2.00 & 52.2 (33.5) & 2.18 & 0.35 & 0.14 & 0.08 & 0.15 & 0.16 \\ 
EB40 & SER3*** & 286.46 & $-$0.34 & - & 250 & 0.03 & 62.2 (29.0) & 1.88 & 0.28 & 0.17 & 0.23 & 0.15 & 0.16 \\ 
EB41 & J1117-6120 & 291.88 & $-$0.50 & 7600 & 2600 & 0.60 & 52.9 (32.3) & 1.26 & 0.27 & 0.07 & 0.24 & 0.17 & 0.11 \\ 
EB42 & SER7*** & 347.15 & 2.36 & - & - & - & 13.2 (7.7) & 0.79 & 0.13 & 0.21 & 0.15 & 0.28 & 0.08 \\ 
EB43 & G4 & 352.57 & 2.11 & 1700 & 2550 & 0.50 & 4.8 (1.5) & 1.82 & 0.26 & 0.27 & 0.48 & - & 0.22 \\ 
EB44 & G2** & 352.81 & 1.34 & 1700 & 2250 & 0.40 & 20.9 (8.7) & 1.08 & 0.21 & 0.16 & 0.36 & 0.57 & 0.13 \\ 
EB45 & G5 & 351.65 & 0.51 & 1700 & 2000 & 0.10 & 28.0 (11.9) & 0.49 & 0.32 & 0.21 & 0.28 & 0.55 & 0.18 \\ 
EB46 & G6 & 353.06 & 1.29 & 1700 & [1000] & 0.10 & 30.1 (11.6) & 0.32 & 0.14 & 0.11 & 0.25 & 0.29 & 0.07 \\ 
EB47 & G8 & 353.16 & 1.05 & 1700 & [1500] & 0.04 & 34.9 (16.6) & 0.68 & 0.34 & 0.22 & 0.28 & 0.24 & 0.20 \\ 
EB48 & G1 & 353.42 & 0.45 & 1700 & 2100 & 0.20 & 56.4 (31.1) & 0.56 & 0.22 & 0.12 & 0.09 & 0.26 & 0.12 \\ 
EB49 & G7 & 354.03 & 0.85 & 1700 & [1000] & 0.10 & 38.0 (20.5) & 0.18 & 0.10 & 0.11 & 0.05 & 0.08 & 0.04 \\ 
EB50 & G3 & 353.30 & 0.08 & 1700 & 2000 & 0.40 & 48.3 (29.9) & 0.85 & 0.29 & 0.09 & 0.24 & 0.05 & 0.19 \\ 
EB51 & HIP 86768 & 18.70 & 11.60 & 737 & [550] & 0.03 & 1.3 (0.4) & 9.39 & 1.45 & 0.98 & 3.45 & 3.44 & 0.94 \\ 
EB52 & Star 1** & 16.99 & 1.77 & 1800 & 2200 & 0.63 & 20.6 (13.9) & 0.65 & 0.18 & 0.18 & 0.22 & 0.25 & 0.12 \\ 
EB53 & M 17-S1* & 15.08 & 0.65 & 1600 & 1000 & 0.03 & 22.7 (6.6) & 0.59 & 0.09 & 0.13 & 0.17 & 0.32 & 0.06 \\ 
EB54 & M 17-S2 & 15.08 & 0.65 & 1600 & [1500] & 0.16 & 22.7 (6.6) & 0.59 & 0.09 & 0.13 & 0.17 & 0.32 & 0.06 \\ 
EB55 & M 17-S3* & 15.08 & 0.65 & 1600 & 2300 & 0.25 & 22.7 (6.6) & 0.59 & 0.09 & 0.13 & 0.17 & 0.32 & 0.06 \\ 
EB56 & BD -14 5040 & 16.89 & $-$1.12 & 1800 & 400 & 0.03 & 111.3 (73.2) & 0.28 & 0.09 & 0.13 & 0.09 & 0.09 & 0.09 \\ 
EB57 & 4U 1907+09 & 43.74 & 0.47 & 4000 & 2900 & 0.70 & 94.2 (63.1) & 0.92 & 0.06 & 0.10 & 0.06 & 0.11 & 0.04 \\ 
EB58 & HIP 98418 & 71.60 & 2.90 & 529.1 & 2545 & 0.24 & 4.1 (3.0) & - & 55.94 & 1.48 & 0.81 & 1.00 & 1.46 \\ 

\hline                                             
\end{tabular}
\tablefoot{* The bow shocks M 17-S1, M 17-S2 and M 17-S3 are closer than 0.1 degree and therefore not resolvable by \hess, the upper limits are calculated for the position of M 17-S2 but valid for all three bow shock candidates. ** The analysis was done for the bow shock coordinates, see text for more details. *** The coordinates listed in \seboss are the apex coordinates of the bow shock, not the star's.  $^a$ Distance uncertainty wrong in \seboss (1900$\pm0.1$\,pc), original paper \citep{Sadakane1985} 1.9$\pm0.2$\,kpc.  
}
\end{sidewaystable*}

\subsection*{Power considerations}
For the bow shocks with known stellar parameters and distance, the kinetic power of the wind can be compared to the upper limits of the radiative power at very-high energies. 
The kinetic power in the wind is given by: 
\begin{eqnarray}
P_\mathrm{wind} = \frac{1}{2} \dot{M} v_\mathrm{wind}^2,
\label{eq:p_wind}
\end{eqnarray}
with the mass-loss rate $\dot{M} $ and wind velocity $ v_\mathrm{wind} $ listed in Table \ref{table:UpperLimits}. The integrated upper limit of the VHE radiative power $P_\mathrm{UL}$ is calculated using the VHE flux upper limits derived in this work (see Sect.\,\ref{sec:results}):
\begin{eqnarray}
P_{\mathrm{UL}} &=& 4 \pi d^2 \int_{E_{\mathrm{min}}}^{E_{\mathrm{max}}} dE (E d\Phi/dE), 
\end{eqnarray}
with the distance $d$ listed in Table\,\ref{table:UpperLimits}. The unknown uncertainties of the distances are treated as a systematic caveat here and are not included in the calculation.    
For this power calculation, the upper limits in the $0.1 -10$\,TeV bin, shown in the last column in Table\,\ref{table:UpperLimits}, are used.   

\begin{figure}
  \includegraphics[width=0.5\textwidth]{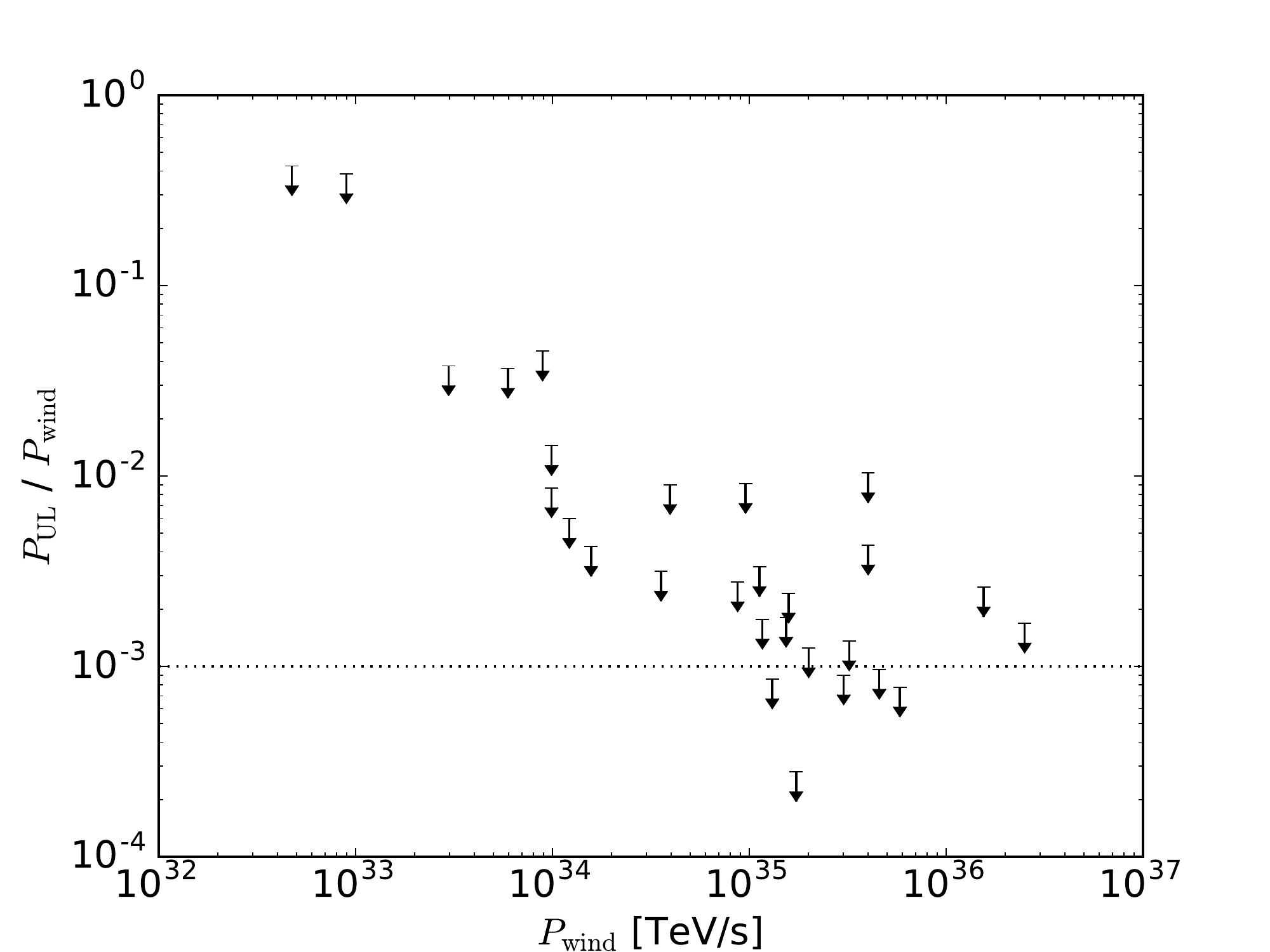}
  \caption{Ratio of power in VHE gamma rays and power in the wind as a function of wind power for the 28 bow shock candidates where the runaway star is identified. The dotted line depicts 0.1\% of the kinetic wind energy. }
  \label{Fig:power}
\end{figure}

Figure\,\ref{Fig:power} shows the ratio of the powers ($P_{\mathrm{UL}} / P_\mathrm{wind} $) as a function of the wind power. We constrain the fraction of wind power that is converted into the production of VHE gamma rays in all cases. In five cases we show that less than 0.1$\%$ of the wind power is potentially converted into the production of VHE gamma rays, while the majority of the limits constrain the ratio of the powers to $<0.1-1\%$. 
One should note that not all of the wind's kinetic power (as given in Eq.\,\ref{eq:p_wind}) is available for particle acceleration: the wind is emitted isotropically, while the bow shock covers only a limited solid angle.

Our systematic population study reveals no evidence for VHE gamma-ray emission from the bow shocks of runaway stars observed in the \hess dataset. Together with the HE gamma-ray upper limits by \cite{Schulz_fermi} and several X-ray upper limits, this challenges the level of predicted non-thermal emission from bow shocks of runaway stars published so far (see Fig.\,\ref{Fig:ul_models} and references therein). 
 
One reason for the non-detection could be that particle acceleration is in general less efficient in bow shocks than in known gamma-ray sources. \cite{terada_bd} concluded that the magnetic fields in the bow shocks of runaway stars might be less turbulent compared to those of pulsar wind nebulae or supernova remnants, where gamma-ray emission is detected in many cases. A lower maximum energy of the accelerated particles or lower photon densities could also explain the non-detections. 

For five bow shocks, we calculate that less than 0.1$\%$ of the kinetic power of the wind is converted into VHE gamma rays originating from relativistically accelerated particles. This is roughly the order of magnitude expected from geometrical considerations. For other astrophysical systems, like e.g. novae \citep{fermi_novae}, the fraction of the total energy in electrons compared to the kinetic energy of the ejected mass is $\sim0.1\%$.

In general, the search for non-thermal emission from bow shocks of runaway stars proves to be a challenge: so far, only one detection of non-thermal radio emission has been reported \citep{benaglia_bd} and upper limits in other radio, X-ray, HE gamma-ray and VHE gamma-ray observations. 
 Our paper presents the first VHE gamma-ray observations of this source class.
 
Our population study shows that none of the already observed stellar bow shocks listed in the \seboss emits VHE gamma rays at a flux level detectable with current imaging atmospheric Cherenkov telescopes. 
\bd could also be unique in this source class as the only bow shock emitting non-thermal radiation. 

The future Cherenkov Telescope Array \citep{CTAHinton2013}, with approximately 10 times better sensitivity than current instruments and improved angular resolution, might be able to detect VHE gamma-ray emission from stellar bow shocks and understand the physics of these objects.

\textbf{Acknowledgements}

The support of the Namibian authorities and of the University of Namibia in facilitating the construction and operation of H.E.S.S. is gratefully acknowledged, as is the support by the German Ministry for Education and Research (BMBF), the Max Planck Society, the German Research Foundation (DFG), the French Ministry for Research, the CNRS-IN2P3 and the Astroparticle Interdisciplinary Programme of the CNRS, the U.K. Science and Technology Facilities Council (STFC), the IPNP of the Charles University, the Czech Science Foundation, the Polish Ministry of Science and Higher Education, the South African Department of Science and Technology and National Research Foundation, the University of Namibia, the Innsbruck University, the Austrian Science Fund (FWF), and the Austrian Federal Ministry for Science, Research and Economy, and by the University of Adelaide and the Australian Research Council. We appreciate the excellent work of the technical support staff in Berlin, Durham, Hamburg, Heidelberg, Palaiseau, Paris, Saclay, and in Namibia in the construction and operation of the equipment. This work benefited from services provided by the H.E.S.S. Virtual Organisation, supported by the national resource providers of the EGI Federation.

This publication makes use of data products from the Wide-field Infrared Survey Explorer, which is a joint project of the University of California, Los Angeles, and the Jet Propulsion Laboratory/California Institute of Technology, funded by the National Aeronautics and Space Administration.

\bibliographystyle{aa}
\bibliography{bowshock_references_v1}

\end{document}